\documentclass[a4paper,10pt,aps,amsfonts,amsmath,nofootinbib,showpacs,byrevtex,prd,reprint]{revtex4-1}   

\usepackage{graphicx}

\newcommand{\cD}{\mathcal{D}}
\newcommand{\cU}{\mathcal{U}}
\newcommand{\cF}{\mathcal{F}}
\newcommand{\cP}{\mathcal{P}}
\newcommand{\be}{\begin{equation}}
\newcommand{\ee}{\end{equation}}
\newcommand{\rk}{\right)}
\newcommand{\lk}{\left(}
\newcommand{\sli}{\sum\limits}

\newcommand{\il}{\int\limits}
\newcommand{\im}{\mathrm{i}}

\newcommand{\vA}{\vec{A}}
\newcommand{\vx}{\vec{x}}
\newcommand{\vy}{\vec{y}}
\newcommand{\vz}{\vec{z}}
\newcommand{\vD}{\vec{D}}
\newcommand{\vk}{\vec{k}}
\newcommand{\vp}{\vec{p}}
\newcommand{\vq}{\vec{q}}

\newcommand{\vQ}{\vec{Q}}

\usepackage{bm} 
\renewcommand*{\vec}[1]{\ensuremath{\bm{\mathrm{#1}}}}
\DeclareMathOperator{\Tr}{Tr}	
\DeclareMathOperator{\Det}{Det}
\newcommand*{\dbar}[1][]{\mathop{\mathrm{d}\mkern-7mu\mathchar'26\mkern-1mu^{#1}}\mkern-4mu}
\renewcommand*{\d}[1][]{\mathop{\mathrm{d}^{#1}}\mkern-3mu}

\begin{document}

\title{Variational approach to Yang-Mills theory at finite temperatures}

\author{Hugo Reinhardt,$^1$ Davide R. Campagnari,$^1$ and Adam P. Szczepaniak$^2$}
\affiliation{$^1$ Institut f\"ur Theoretische Physik, Auf der Morgenstelle 14, D-72076 T\"ubingen, Germany  \\
$^2$  Physics Department and Center for Exploration of Energy and Matter, Indiana University, Bloomington, IN 47403, USA} 

\date{\today}

\begin{abstract}
We study the finite-temperature phase of a gluon ensemble in a variational approximation
to QCD in the Coulomb gauge. We derive and numerically solve the underlying Dyson-Schwinger
equations up to one-loop order. 
Assuming the subcritical solution at $T=0$, we find 
a sharp transition in the infrared value of the gluon energy at a critical temperature.
\end{abstract}

\maketitle

\section{Introduction}
\label{intro} 
  
The determination of the phases of hadronic matter plays a major role in understanding the mechanisms
of confinement and dynamical symmetry breaking in Quantum Chromodynamics (QCD). 
The methods available to investigate these phase transitions are the following:
There are lattice simulations, which have
been successful in mapping out the deconfinement transition of the QCD phase diagram as
a function of temperature for near zero chemical potential~\cite{ft-lattice1,ft-lattice2,ft-lattice3,ft-lattice4,ft-lattice5},
and phenomenological models, that in addition can cover the high-density regime \cite{Reinhardt:1987da,pnjl1,pnjl2,pnjl3,pnjl4}.
Finally, in the asymptotically large temperature or density limit, due to asymptotic freedom,
the weak interactions between quarks and gluons are expected to determine the properties
of the quark-gluon plasma~\cite{hard1,hard2,hard3,hard4,hard5}.

In this paper we investigate the thermal properties of the low-density phase using a set of
tools that bridge QCD and phenomenology. In particular, we formulate the problem in the
physical,
Coulomb gauge, canonical Hamiltonian framework of the pure gauge theory. While
there have been numerous studies of QCD based on Dyson-Schwinger resummation 
techniques~\cite{Maas:2004se,ftsd1,ftsd2,ftsd3}, renormalization group flow equations \cite{Pawlowski:2010ht},
and lattice simulations \cite{Cucchieri:2007ta,Bornyakov:2010nc,Cucchieri:2011di}
in covariant gauges, the few that exist in physical gauges are rather loosely related to
the underlying QCD interactions~\cite{pg1,pg2,pg3,pg4,pg6,pg7}, with a recent attempt
at a self-consistent calculation at finite density~\cite{Guo:2009ma}.\footnote{At zero temperature
and density, Dyson-Schwinger studies in Coulomb gauge have been performed in
Refs.~\cite{Watson:2006yq,Popovici:2010mb,Watson:2010cn}.}

The advantages of
physical gauges for phenomenology and for developing physical intuition are clear, and we summarize them here. 
The degrees of freedom of the pure Yang-Mills (YM) theory are transverse gluons, and thermal
excitations connect color singlet states of arbitrary number of gluons.
Transverse gluons are expected to be
effective only at high temperatures, while at low temperatures it would be more effective
to compute the partition function in terms of 
the ground state glueballs~\cite{Szczepaniak:1995cw,Szczepaniak:2003mr}.
The underlying interactions in Coulomb gauge are dominated by the instantaneous Coulomb
potential acting between color charges. In the non-Abelian theory, the potential not only
couples charges but it also depends on the gluon distribution of the state in which it is
calculated.  At zero temperature, in the vacuum state 
this distribution is such that the Coulomb potential becomes confining, i.e., proportional
to the distance $R$ between the external color charges, $V(R) = \sigma_{\textsc{c}} R$~\cite{Zwanziger:2002sh,Greensite:2004ke}.
Using various approximate, variational models for the ground state YM wave functional,
it has been possible to obtain a potential which is confining \cite{Epple:2006hv} or
almost confining, i.e., $V(R) \propto R^{1 - \epsilon} $ with 
   $\epsilon  \approx O(10\%)$~\cite{Szczepaniak:2001rg,Feuchter:2004mk,Epple:2007ut}. The Coulomb string tension 
$\sigma_{\textsc{c}}$ is larger than the string tension computed from the temporal
Wilson loop. This is because the Coulomb potential represents the energy of a static quark-antiquark pair
 submersed in the QCD vacuum, while the Wilson loop measures the energy of the exact  $Q{\bar Q}$ state in which the 
gluon distribution is squeezed by closed vortex lines. 
Since the Coulomb potential is an instantaneous observable, one might expect that it remains
confining even in the high-temperature limit~\cite{Greensite:2004ke}:
At high temperatures the integration over transverse  
fields becomes even less restricted than in the vacuum, and, according to the Gribov-Zwanziger
confinement scenario~\cite{Gribov:1977wm,Zwanziger:1993dh}, Coulomb confinement
originates from large field configurations near the the Gribov horizon. 
   
In the following, we investigate the finite temperature properties of Coulomb gauge Yang-Mills theory
with focus on the aspects of deconfinement at finite temperature. We extend the variational
approach of Refs.~\cite{Feuchter:2004mk,Epple:2006hv,Epple:2007ut} to finite temperature.
In particular, the variational Gaussian ansatz for the vacuum wave functional is extended
to include single particle, quasi-gluon excitations. In Sections~\ref{GS} and~\ref{CP} we
present the general setting for the finite temperature, canonical Coulomb gauge problem.
In Section~\ref{VP} we discuss the details of the variational approximation. In Section~\ref{NR}
we give details of the numerical computations and results. Our summary
and outlook are given in Section~\ref{SO}.

\section{\label{GS}Hamiltonian approach at finite temperatures}

After resolving Gauss's law in Coulomb gauge, the Yang-Mills Hamiltonian reads
\begin{equation}\label{398-G1}
\begin{split}
H_{\textsc{ym}} &= \frac{1}{2} \int \d[3] x \left( J^{-1}[\vec{A}] \,{\vec{\Pi}} J[\vec{A}] \,\vec{\Pi}
+ \vec{B}^2 \right) + H_\textsc{c} \\
&\equiv  H_K + H_B + H_\textsc{c} ,
\end{split}
\end{equation}
\begin{equation}\label{398-G2}
H_\textsc{c} = \frac{g^2}{2} \int \d[3]x \d[3]y \: J^{- 1} [\vA] \, \rho^a (\vx) \, J [\vA] \, F^{ab}_A (\vx,\vy) \, \rho^b (\vy) ,
\end{equation}
where $\Pi^a(\vx) = -\im \delta/\delta A^a(\vx)$ is the canonical momentum (electric
field) operator, and
\be
\label{404}
J[\vec{A}] = \Det (- \vec{D} \vec{\nabla})
\ee
is the Faddeev-Popov determinant with
\be
\label{ha-176}
\vD = \vec{\nabla} + g \hat{\vA} , \qquad
\hat{\vA}{}^{ab} = \hat{T}_c \vA^c , \qquad
( \hat{T}_c )^{ab} = f^{acb}
\ee
being the covariant derivative in the adjoint representation. Furthermore,
\be
\label{409-G3}
\rho^a (\vx) = - f^{abc} \vec{A}^b \cdot \vec{\Pi}^c
\ee
is the color charge density of the gluons and
\be
\label{414-G4}
F^{ab}_A(\vx, \vy) = \langle \vx, a \rvert (- \vec{D} \vec{\nabla})^{- 1}
\, (- \vec{\nabla}^2)\, (- \vec{D} \vec{\nabla})^{- 1} \lvert \vy, b \rangle
\ee
is the so-called Coulomb kernel. Its vacuum
expectation value $\langle F^{ab}_A(\vx, \vy)\rangle$ represents the static non-Abelian
color Coulomb potential.

The gauge fixed Hamiltonian
Eq.~(\ref{398-G1}) is highly non-local due to Coulomb kernel $F_A(\vx,\vy)$,
Eq.~(\ref{414-G4}), and due to the Faddeev-Popov determinant $J[\vA]$, Eq.~(\ref{404}).
In addition, the latter also occurs in the functional integration measure of
the scalar product of the Coulomb gauge wave functionals
\be
\label{419-G5}
\langle \psi_1 | O | \psi_2 \rangle = \int D A \, J [\vec{A}] \, \psi^*_1[\vec{A}] \, O \, \psi_2 [\vec{A}] .
\ee
In Ref.~\cite{Feuchter:2004mk} the Yang-Mills Schr\"odinger equation was solved by the
variational principle using the following ansatz for the vacuum wave functional
\be\label{201}
\begin{split}
\langle A | 0 \rangle &= \frac{1}{\sqrt{J[\vec{A}]}} \, \langle A | \tilde{0} \rangle , \\
\langle A | \tilde{0} \rangle &= \mathcal{N} \exp \left( - \frac{1}{2}
\int \dbar{k} \: A (- \vk) \omega (\vk) A (\vk) \right) ,
\end{split}
\ee
where 
\be
\label{226}
\dbar{k} = \frac{\d[3]k}{(2 \pi)^3 } .
\ee
The pre-exponential factor removes the Faddev-Popov determinant from the scalar product
Eq.~(\ref{419-G5}). The kernel $\omega(\vk)$ was determined by minimizing the vacuum
energy $\langle H_{\textsc{ym}} \rangle$, which yields an $\omega(\vk)$ which can be
well fitted by Gribov's formula
\be
\label{208}
\omega (\vk) = \sqrt{\vk^2 + \frac{M^4}{\vk^2}} \, 
\ee
and which is in satisfactory agreement with the lattice data \cite{Burgio:2008jr}, with
$M \approx 860$~MeV.

The present paper is devoted to study Yang-Mills theory at finite temperatures, which is
defined by the  density operator
\be\label{218-3}
{\cD} = Z^{- 1} \exp (- \beta H_{\textsc{ym}}),
\ee
where $\beta = 1/T$ is the inverse temperature and
\be\label{224-4}
Z = \Tr \mathrm{e}^{- \beta H_{\textsc{ym}}}
\ee
is the partition function.

To calculate the trace in the thermal averages
\be
\label{233-5}
\langle O \rangle = \Tr (O {\cD} )
\ee
we need a suitable basis in the gluonic Fock space, which we choose as follows: Let $a^a_i(\vk)$ be the operator
which annihilates the vacuum state $| \tilde{0} \rangle$ [Eq.~(\ref{201})], i.e.,
\be\label{239-6}
a^a_i (\vk) | \tilde{0} \rangle = 0 .
\ee
Then a complete basis in the gluonic Fock space is given by
\be\label{244-7}
\bigl\{ | \tilde{n} \rangle \bigr\} = \bigl\{ | \tilde{0} \rangle , \:
a^{a\dagger}_i (\vk) | \tilde{0} \rangle , \: a^{a\dagger}_i (\vk) a^{b\dagger}_j (\vq) | \tilde{0} \rangle , \: \dots \bigr\} .
\ee
Following Ref.~\cite{Feuchter:2004mk} we choose the basis states of the gluonic Fock space in the form (cf.~Eq.~(\ref{201})) 
\be
\label{250-8}
\bigl\{ | n \rangle \bigr\} = \bigl\{ J^{-1/2}[\vec{A}] | \tilde{n} \rangle \bigr\} .
\ee
The thermal expectation value Eq.~(\ref{233-5}) can then be expressed as
\be
\label{255-9}
\langle O \rangle = \mathrm{\tilde{T}r} ( \tilde{\cD} \tilde{O} ) ,
\ee
where the operation `$\sim$' is defined by
\be
\label{260-10}
\tilde{O} = J^{1/2}[\vec{A}] O J^{- 1/2}[\vec{A}] , 
\ee
and `$\mathrm{\tilde{T}r}$' means that the trace is evaluated in the basis of the states
$\left\{ | \tilde{n} \rangle \right\}$  [Eq.~(\ref{244-7})]. The transformed density
operator reads explicitly
\be
\label{266-11}
\tilde{\cD} = Z^{-1} \exp( - \beta \tilde{H}_{\textsc{ym}} ) , \qquad
Z = \mathrm{\tilde{T}r} \, \mathrm{e}^{- \beta \tilde{H}_{\textsc{ym}}} .
\ee
This operator is too difficult to handle in semi-analytical calculations. In analogy to the zero-temperature
case, where the Gaussian vacuum wave functional Eq.~(\ref{201}) was assumed, we will replace the exact (transformed according
to Eq.~(\ref{260-10})) Yang-Mills Hamiltonian $\tilde{H}_{\textsc{ym}}$ by a single-particle operator
\be
\label{273-12}
\tilde{h} =  \int \dbar{k} \: \Omega(\vk) \, a^{b\dagger}_i(\vk) \, a^b_i(\vk) ,
\ee
where the kernel $\Omega(\vk)$ will be determined by minimizing the free energy
\be
\label{278-13}
\cF = \langle H_{\textsc{ym}} \rangle - TS .
\ee
Here $S$ is the entropy, which is defined by
\be
\label{283-14}
S = - \mathrm{\tilde{T}r} \tilde{\cD} \ln \tilde{\cD} .
\ee
By straightforward manipulations the expression for the entropy Eq.~(\ref{283-14}) can be cast into the form
\be
\label{325-18}
S = \ln Z - \beta \frac{\partial \ln Z}{\partial \beta} \, ,
\ee
which will be convenient in later calculations.


\section{\label{CP}Color projection}

\subsection{\label{subsec:proj}Exact projection}

By definition the trace in the thermal averages Eq.~\eqref{255-9} should be taken
in the physical Hilbert space. Before gauge fixing
the physical Hilbert space is given by all gauge invariant states. After
resolving Gauss' law in Coulomb gauge, the
physical Hilbert space is defined by a complete set of wave functionals
of the transversal gauge field that are invariant
under global gauge transformations (the latter is not fixed by the
Coulomb gauge condition). These states are annihilated by the
total color charge operator\footnote{Also in the functional integral
formulation after fixing to Coulomb gauge a
careful treatment of the zero modes of the Faddeev-Popov operator
related to the global gauge transformations constrains
the ensemble of transversal gauge fields to those with vanishing total
color charge \cite{Reinhardt:2008pr}.}.
\be
\label{303}
Q^a = \int \d[3]x \: \rho^a (\vx) .
\ee
However, an individual basis state of the set Eq.~(\ref{244-7}) will, in general, carry a non-zero color charge,
and the use of the basis Eq.~(\ref{250-8}) will lead to a colored statistical ensemble. Therefore we
project these states onto color singlet states using the projector 
\be
\label{38}
\mathcal{P} = \int \d\mu(\vec{\theta}) \exp \left[ i \theta_a Q^a \right],
\ee
where $\d\mu(\vec{\theta})$ denotes the Haar
measure of the gauge group parametrized in terms of the color angles $\theta_a$. The
thermal average projected onto zero-color states reads
\be
\label{305-15}
\langle O \rangle = \mathrm{\tilde{T}r} \bigl( \tilde{O} \tilde{\cD} \cP \bigr) ,
\ee
where $\tilde{\cD}$ is given by Eq.~(\ref{266-11}) with $\tilde{H}_{\textsc{ym}}$
replaced by $\tilde{h}$ [Eq.~(\ref{273-12})]:
\be
\label{310-16}
\tilde{\cD} = Z^{- 1} \mathrm{e}^{- \beta \tilde{h}} , \qquad
Z = \mathrm{\tilde{T}r} \bigl( \mathrm{e}^{- \beta \tilde{h}} {\cP} \bigr) .
\ee
The density operator $\tilde{\cD}$ [Eq.~(\ref{310-16})] is color singlet and hence
commutes with the total color charge operator $Q^a$ [Eq.~(\ref{303})], which in terms
of the creation and annihilation operators reads
\be
\label{316-17}
Q^a = \im f^{abc} \int \dbar{k} \:  a^{b\dagger}_i(\vk) \, a^c_i(\vk) .
\ee
With the explicit form of the projector Eq.~(\ref{38}) we have
\be
\label{337}
\tilde{\cD} {\cP} = \int \d\mu(\vec{\theta}) \: {\cD}_{\vec{\theta}} \, ,
\ee
where
\be
\label{ha-496}
\cD_{\vec{\theta}} = \mathrm{e}^{\im \vec{\theta} \cdot \vQ} \tilde{\cD} =
\mathrm{e}^{- \beta \tilde{h} + \im \vec{\theta} \cdot \vQ}
\ee
is the density operator in the presence of an external color field $(- \im \theta_a Q^a / \beta )$,
i.e., for fixed color angle $\theta_a$. Due to the presence of the external color field
$( - \im \theta_a Q^a / \beta )$, this density matrix is non-diagonal in color space.
However, since the total charge operator $Q_a$ is hermitian and $\tilde{h}$ is color
singlet, we can diagonalize ${\cD}_{\vec{\theta}}$. For simplicity, we consider the
gauge group $SU(2)$. Then we may write
\be
\label{52}
\cD_{\vec{\theta}}  =  \cU^\dagger (\Hat{\bm{\theta}}) \cD_\theta \, {\cU} (\Hat{\bm{\theta}}) ,
\ee
where ${\cU}(\Hat{\bm{\theta}})$ lives in the coset $SU (2) / U (1)$ and 
\be
\label{ha-537}
\cD_\theta =  \exp \bigl( - \beta \tilde{h} + \im \theta Q^3 \bigr)
\ee
lives in the Abelian subgroup. In a parametrization of the gauge group $SU(2)$ 
corresponding to the coset decomposition Eq.~(\ref{52}) the Haar measure reads
\be\label{499-55}
\begin{split}
\int_{S^3} \d\mu(\vec{\theta}) &= \int \d\mu(\theta)  \int_{S^2} \d\mu(\Hat{\bm{\theta}}) , \\
\int \d\mu(\theta) &= \frac{1}{\pi} \int^\pi_{-\pi} \d\theta \: \sin^2\frac{\theta}{2} \, ,
\end{split}
\ee
where $\d\mu(\Hat{\bm{\theta}})$ denotes the measure for the integration over the
coset's $SU (2) / U (1) \simeq S^2$ degrees of freedom. 

In the thermal averages [Eq.~(\ref{305-15})] of colorless operators $O$ the unitary matrix
${\cU} (\Hat{\bm{\theta}})$ drops out. Since the density matrix $\cD_\theta$ does not
depend on the coset degrees of freedom $\Hat{\bm{\theta}}$, the corresponding integral
can then be trivially carried out
\be
\label{56}
\int_{S^2} \d \mu (\Hat{\bm{\theta}}) = 4 \pi ,
\ee
and we obtain for the projected thermal averages
\be
\label{53}
\langle O \rangle = \frac{1}{Z} \int \d\mu(\theta) \: Z (\theta) \, \langle O \rangle_\theta \, , \quad
Z = \int \d\mu(\theta) \: Z (\theta) ,
\ee
where
\be
\label{54}
\langle O \rangle_\theta = \frac{1}{Z (\theta)} \, \mathrm{\tilde{T}r} (\mathcal{D}_\theta \tilde{O}),
\qquad Z (\theta) = \mathrm{\tilde{T}r} {\cD}_\theta
\ee
denotes the thermal expectation value for a fixed color angle $\theta$. 

Furthermore, it is also convenient to use the basis in color space in which 
$(\hat{T}_3)^{ab} = \varepsilon^{a3b}$ is diagonal. In this basis we have
\be
\label{ha-593}
Q^3 = \sli_{\alpha = 0, \pm 1} \alpha \int \dbar{k} \: a^{\alpha\dagger}_i (\vk) \, a^\alpha_i (\vk) ,
\ee
and the density operator $\cD_\theta$ [Eq.~(\ref{ha-537})] becomes 
\be
\label{62}
{\cD}_\theta = \exp \biggl[ - \beta \sli_{\alpha,i} \int \dbar{k} \:
\varepsilon^\alpha(\vk,\theta) \, a^{\alpha\dagger}_i(\vk) \, a^\alpha_i(\vk) \biggr],
\ee
where
\be
\label{63}
\beta \varepsilon^\alpha (\vk, \theta) = \beta \Omega (\vk) - i \theta \alpha .
\ee
Since $\cD_\theta$ is the (exponent of a) single particle operator, the thermal expectation values 
$\langle{\dots}\rangle_\theta$ can be evaluated using Wick's theorem. In the standard fashion one finds for the
partition function
\be
\label{73}
Z (\theta) = \exp \biggl\{ 2 V \sli_\alpha \int \dbar{k} \ln [1 + n_\alpha (\vk, \theta)] \biggr\} \, ,
\ee
where $V$ is the volume of ordinary space and $2 = t_{ii} (\vk)$ is the number of independent polarization 
degrees of freedom in three
dimensions.
Furthermore,
\be
\label{65}
n_\alpha (\vk, \theta) = \lk \mathrm{e}^{\beta \varepsilon^\alpha (\vk, \theta)} - 1 \rk^{- 1}
\ee
are the finite-temperature Bose occupation numbers. The basic contraction is obtained as
\be
\label{64}
\langle a^{\alpha\dagger}_i (\vk) a^\beta_j (\vq) \rangle =
\delta^{\alpha \beta} t_{ij}(\vk) \, (2\pi)^3 \delta(\vk-\vq) \, n_\alpha (\vk, \theta) .
\ee
Expressing the gauge field in terms of the creation and annihilation operators one finds
from Eq.~(\ref{64}) for the gluon propagator
\be\label{465-21}
\langle A^\alpha_i (\vp) A^\beta_j (\vq) \rangle =
\delta^{\alpha \beta} \, t_{ij}(\vp) \, (2\pi)^3 \delta(\vp + \vq) \frac{1 + 2 n_\alpha (\vp,\theta)}{2 \omega (\vp)} \, .
\ee
With these relations it is straightforward to calculate the thermal expectation value of
the Hamiltonian using the same approximation as at zero temperature in Ref.~\cite{Feuchter:2004mk},
i.e., assuming a bare ghost-gluon vertex and calculating the energy up to two loops.

To work out the effect of the color projection on the energy, let us for the moment ignore
the Faddeev-Popov determinant in the Hamiltonian. We will later fully include $J[\vec{A}]$.
Using the explicit form of the thermal gluon propagator, Eq.~(\ref{465-21}), and the same
approximation as in Ref.~\cite{Feuchter:2004mk} but putting $J[\vec{A}] = 1$, one finds
for the various pieces of the energy
\begin{subequations}\label{projvev}
\be
\label{465-1}
\langle H_K \rangle_\theta = \frac{V}{4} \int \dbar{q} \: \omega(\vq)
\bigl[ 3+2\sum_\alpha n_\alpha(\vq) \bigr],
\ee
\begin{multline}\label{467-2}
\langle H_B \rangle_\theta =
\frac{V}{2} \int \dbar{q} \: \frac{\vq^2}{\omega(\vq)} \bigl[ 3+2\sum_\alpha n_\alpha(\vq) \bigr] \\
+V \frac{g^2 N_\mathrm{c}}{16} \int \dbar{q} \dbar{p} \frac{3-(\hat{\vq}\cdot\hat{\vp})^2}{\omega(\vq) \, \omega(\vp)} \\
\times \biggl[ 3 + 2\sum_\alpha\bigl( n_\alpha(\vp) + n_\alpha(\vq) \bigr) + 2 \sum_{\alpha,\beta} n_\alpha(\vp) \, n_\beta(\vq) \\
- \sum_\alpha n_\alpha(\vp) \bigl( n_\alpha(\vq) + n_{-\alpha}(\vq) \bigr) \biggr],
\end{multline}
\begin{widetext}
\begin{multline}\label{501-2}
\langle H_{\textsc{c}} \rangle_\theta =  \frac{g^2N_\mathrm{c}}{8} V \int \dbar{q} \, \dbar{p} \: \bigl[1+(\hat{\vq}\cdot\hat{\vp})^2\bigr] \,
F (\vq-\vp) \\
 \biggl\{ \frac{\omega(\vq)}{\omega (\vp)} \biggl[ 3 + 2 \sli_\alpha \bigl( n_\alpha (\vq) + n_\alpha (\vp) \bigr) +
2 \Bigl( \sli_\alpha n_\alpha(\vq) \Bigr) \Bigl( \sli_\beta n_\beta (\vp) \Bigr) \\
 - \sli_\alpha n_\alpha (\vq) \bigl( n_\alpha (\vp) + n_{-\alpha}(\vp)\bigl) \bigg] 
-3 + \sli_\alpha n_\alpha (\vp) \bigl( n_\alpha (\vq) - n_{-\alpha}(\vq) \bigr) \biggr\} \\
 + \frac{g^2 N_\mathrm{c}}{8} \, V \cdot 2 \cdot F (0) \int \dbar{p} \dbar{q} \sum_{\alpha} n_\alpha (\vp)
\bigl( n_\alpha (\vq) - n_{- \alpha} (\vq) \bigr) .
\end{multline}
\end{widetext}
\end{subequations}
To simplify the notation, we have omitted the $\theta$-de\-pen\-dence of the occupation
numbers Eq.~\eqref{65}. In Eq.~\eqref{501-2}
\be
\label{485-3}
F (\vx, \vy) = \langle F_A(\vx, \vy) \rangle
\ee
is the non-Abelian color Coulomb potential. This quantity is known from the lattice and also
from continuum studies \cite{Greensite:2004ke,Epple:2006hv} to have the infrared behavior
\be
\label{491-4}
F(\vk \to 0) \sim 1 /k^4 .
\ee
Accordingly, the integrand in Eq.~(\ref{501-2}) becomes divergent for $\vp = \vq$.
Furthermore, the last term in Eq.~(\ref{501-2}) is manifestly divergent. However, one
easily shows that these divergent terms disappear after color projection. For this purpose,
we note that if one replaces in the Coulomb Hamiltonian $H_{\textsc{c}}$ [Eq.~(\ref{398-G2})]
the Coulomb kernel $F_A$ by
\be
\label{499-5}
\frac{g^2}{2} F^{ab}_A (\vx, \vy) \to \delta^{ab} \delta(\vx-\vy) ,
\ee
the Coulomb Hamiltonian becomes the square of the total charge
\be
\label{504-6}
H_{\textsc{c}} \to Q^a Q^a .
\ee
This equivalence holds even when the Faddeev-Popov determinant is included,
since $J[\vA]$, being invariant under global color rotations,
commutes with the total color charge operator $Q^a$. In momentum
space the replacement Eq.~(\ref{499-5}) corresponds to
\be
\label{511-7}
\frac{g^2}{2} F(\vk) \to (2 \pi)^3 \delta (\vk) .
\ee
It follows that the singular $\vp = \vq$ contributions to the double integral in $\langle H_{\textsc{c}} \rangle_\theta$
[Eq.~(\ref{501-2})] are proportional to $\langle Q^a Q^a \rangle_\theta$. However, this quantity has to vanish
after color projection
\be
\label{518-7}
{\langle Q^a Q^a \rangle} = \frac{1}{Z}\int \d \mu (\theta) \:
Z(\theta) \, \langle Q^a Q^a \rangle_\theta = 0 .
\ee
Therefore, the singular contributions that occur from the $\vp = \vq$ part of the integrand
vanish after color projection. We can explicitly eliminate these singularities by replacing
the Coulomb potential $F(\vk)$ by 
\begin{align}
\bar{F} (\vk) &= F (\vk) - F (0) V^{- 1} (2 \pi)^3 \delta (\vk) \nonumber \\
& = F (\vk) \bigl( 1 - V^{- 1} (2 \pi)^3 \delta (\vk ) \bigr) . \label{525-8}
\end{align}
This replacement will, in particular, remove the last term of Eq.~(\ref{501-2}). The
kernel $\bar{F}$ and thus the color-projected Coulomb energy $\langle H_\textsc{c} \rangle_\theta$
is invariant with respect to a shift of the Coulomb kernel by a constant
\be
\label{533-9}
F (\vx, \vy) \to F (\vx, \vy) + C .
\ee
This shift implies in momentum space
\be
\label{538-10}
F (\vk) \to F (\vk) + C (2 \pi)^3 \delta (\vk) ,
\ee 
which obviously leaves $\bar{F} (\vk)$ [Eq.~(\ref{525-8})] unchanged.

\subsection{Color projection in the thermodynamic limit}

The partition function Eq.~(\ref{73}) depends via the finite-temperature occupation
numbers $n_\alpha (\vk)$ [Eq.~\eqref{65}] on the color angle $\theta$. The $\theta$-dependence can be
explicitly separated yielding
\be
\label{74}
Z (\theta) = Z (0) \exp (- V f (\theta)) ,
\ee
where
\be
\label{75}
Z (0) = \exp \left\{ 2 (N_{\mathrm{c}}^2-1) V \int \dbar{k} \ln [1 + n (\vk)] \right\}
\ee
is the partition function for vanishing ``external'' color field $(\theta = 0)$,
with
\be
\label{372}
n (\vk) = n_\alpha (\vk) |_{\theta = 0} = n_{\alpha = 0} (\vk) = \lk \mathrm{e}^{\beta \Omega (\vk)} - 1 \rk^{- 1}
\ee
being the corresponding thermal occupation numbers. The $\theta$-dependence is entirely
contained in the exponent of Eq.~(\ref{74}), which is given by
\be
\label{76}
f (\theta) = 2 \int \dbar{k} \ln \bigl[ 1 + 2 (1 - \cos\theta) \, n(\vk) \bigl( 1 + n (\vk)\bigr) \bigr] .
\ee
Note that the partition function is an even function in $\theta$. This property holds
for the expectation value $\langle O \rangle_\theta$ of any color singlet operator $O$.

Consider now the total partition function $Z$, Eq.~(\ref{53}). In the integration domain
$\theta \in [- \pi, \pi]$ the function $f(\theta)$ [Eq.~(\ref{76})] takes its minimum
at $\theta = 0$, where it vanishes
\be
\label{78}
f (\theta = 0) = 0  .
\ee
Due to the presence of the volume factor $V$, in the thermodynamic limit $V \to \infty$
only small $\theta$ values contribute to the integral Eq.~(\ref{76}). Therefore it suffices
to expand the function $f (\theta)$ to leading order in $\theta$ yielding\footnote{The same expansion was used 
in Ref.~\cite{LeYaouanc:1987ct} for the quark partition function.} 
\be
\label{79}
Z (\theta)  = Z (0) \, \mathrm{e}^{- \frac{1}{2} C \theta^2} ,
\ee
where
\be
\label{80}
C = V f'' (0) , \qquad f'' (0) = 2 \int \dbar{k} \: n(\vk) \bigl(1 + n (\vk)\bigr) .
\ee
With this representation for $Z (\theta)$, the thermal expectation value Eq.~(\ref{53})
of an observable $O$ becomes
\be
\label{81}
\langle O \rangle = \frac{Z (0)}{Z} \int \d\mu(\theta) \: \langle O \rangle_\theta \, \mathrm{e}^{- \frac{1}{2} C \theta^2}
\ee
with
\be
\label{82}
Z = Z (0) \int \d \mu (\theta) \: \mathrm{e}^{- \frac{1}{2} C \theta^2} .
\ee
Due to the presence of the Gaussian, only small $\theta$ values contribute significantly to the integrals. 
Therefore it suffices to expand $\langle O \rangle_\theta$ up to leading order in $\theta$
\be
\label{86}
\langle O \rangle_\theta = \langle O \rangle_{\theta = 0}  + O^{(2)} \theta^2 + \cdots
\ee
Defining
\be
\label{83}
I_n = \frac{1}{Z (0)} \int \d\mu(\theta) \: Z(\theta) \, \theta^{2 (n - 1)}
\ee
and using
\be
\label{84}
Z = Z (0) I_1 \, ,
\ee
we obtain
\be
\label{85}
\langle O \rangle = \langle O \rangle_{\theta = 0} + \frac{I_2}{I_1} O^{(2)} .
\ee
Along the same lines, we can also expand the integration measure $\d\mu (\theta)$ [Eq.~(\ref{499-55})]
to leading order in $\theta$
and put the upper integration limit to $\infty$. This yields for the integrals Eq.~(\ref{83})
\be
\label{88}
I_n = \frac{1}{2 \pi} \il^\infty_0 \d \theta \: \theta^{2 n} \, \mathrm{e}^{- \frac{1}{2} C \theta^2} = \frac{1}{\sqrt{8 \pi C}} \frac{(2 n - 1)!!}{ C^n} \, .
\ee
Since $I_2 / I_1 \sim 1/V$ in the thermodynamic limit $V \to \infty$, the second term
in Eq.~(\ref{85}) can be omitted and we find
\be
\label{89}
\langle O \rangle = \langle O \rangle_{\theta = 0} \, .
\ee
This shows that in leading order in the thermodynamic limit the effect of the color projection can be ignored.
In the following we will skip the subscript $\theta = 0$ and $\langle O \rangle$ means $\langle O \rangle_{\theta = 0}$,
which is the unprojected thermal average. 

To include the Faddeev-Popov determinant we use the representation \cite{Reinhardt:2004mm}
\be
\label{735-XX}
J[\vA] = \exp \lk - \frac{1}{2} \int \dbar{k} \: A (-\vk) \chi (\vk) A (\vk) \rk ,
\ee
where 
\be
\label{490-25}
\chi (\vp) =
\frac{N_\mathrm{c}}{4} \int \dbar{p} \lk 1 - \lk \hat{\vp} \cdot \hat{\vq} \rk^2 \rk \frac{d(\vp-\vq) \, d(\vq)}{\lk \vp - \vq \rk^2}
\ee
is the ghost loop (curvature) and $d (\vp)$ is the ghost form factor defined by
\be
\label{496-26}
\langle (- D \partial)^{- 1} \rangle = \frac{1}{g} \frac{d (- \Delta)}{(- \Delta)} \, ,
\ee
which satisfies the following Dyson-Schwinger equation
\be\label{501-27}
\begin{split}
d^{- 1} (\vp) &= \frac{1}{g} - I_d (\vp) , \\
I_d (\vp) &= \frac{N_\mathrm{c}}{2} \int \dbar{q} \bigl[ 1 - (\hat{\vp}\cdot\hat{\vq})^2 \bigr]
\frac{d ( \vp - \vq )}{(\vp - \vq)^2} \frac{1+2n(\vq)}{\omega(\vq)} ,
\end{split}
\ee
where a bare ghost-gluon vertex has been assumed.
This equation differs from the zero-temperature case only by the replacement of the gluon
 propagator by its finite temperature counterpart Eq.~(\ref{465-21}). The representation
Eq.~(\ref{735-XX}) is valid up to two loops in the energy, which is the order considered
in the present paper.

With the inclusion of the Faddeev-Popov determinant, the thermal expectation value of the
Hamiltonian $\langle H_{\textsc{ym}} \rangle$ given by Eqs.~\eqref{projvev} simplifies
for $\theta = 0$ to
\be
\label{477-23}
\langle H_{\textsc{ym}} \rangle = \lk N^2_\mathrm{c} - 1 \rk \cdot 2 \cdot V \ e , \qquad e = e_K + e_B + e_\textsc{c} \, ,
\ee
where
\begin{subequations}\label{482-24}
\be
e_K = \frac{1}{4} \int \dbar{q} \left\{ \left[ \omega^2 (\vq) + \chi^2 (\vq)  \right]
 \frac{1 + 2 n(\vq)}{\omega(\vq)} - 2 \chi(\vq) \right\}
\ee
\begin{widetext}
\begin{align}
e_B ={}&
\frac{1}{4} \int \dbar{q} \: \vq^2 \frac{1 + 2n(\vq)}{\omega(\vq)}
+ \frac{g^2 N_\mathrm{c}}{32} \int \dbar{p} \, \dbar{q} \bigl[ 3 - (\hat{\vp}\cdot\hat{\vq})^2 \bigr]
\frac{1 + 2n (\vp)}{ \omega (\vp)} \frac{1 + 2n (\vq)}{\omega (\vq)},  \\
e_{\textsc{c}} ={}& \frac{g^2 N_c}{16} \int \dbar{p} \, \dbar{q} \bigl[ 1 + (\hat{\vp}\cdot\hat{\vq})^2 \bigr]
\frac{\bar{F} (\vp - \vq)}{\omega (\vp) \, \omega (\vq)} 
\Bigl\{ \bigl[ \omega^2 (\vp) + \chi^2 (\vp) - \chi (\vp) \chi (\vq) \bigr]
\bigl(1 + 2n (\vp) \bigr) \bigl(1 + 2n (\vq)\bigr) \nonumber \\
&- \omega(\vp) \, \omega(\vq) + 2 \chi(\vp)
\bigl[ \omega(\vq) \bigl(1 + 2n(\vp)\bigr) - \omega(\vp) \bigl((1 + 2n(\vq)\bigr) \bigr] \Bigr\}
\end{align}
\end{widetext}
\end{subequations}
are the energy densities per degree of freedom.


\section{\label{VP}Finite temperature variational principle}

Our ansatz for the density operator [Eqs.~(\ref{273-12}) and (\ref{310-16})] contains a
so far arbitrary kernel $\Omega (\vk)$, which we determine now by minimizing the free energy
${\cF}$, Eq.~(\ref{278-13}). Instead of varying ${\cF}$ with respect to $\Omega (\vk)$,
it is more convenient to take the variation with respect to the finite-temperature
occupation number $n (\vk)$ [Eq.~(\ref{372})], which is a monotonic function of
$\Omega (\vk)$ for $\Omega (\vk) > 0$. Variation of ${\cF}$ with respect to $n(\vk)$ yields
\be
\label{518-28}
\Omega (\vk) = \frac{\delta e [n]}{\delta n (\vk)} \, ,
\ee
which identifies $\Omega (\vk)$ as the quasi-gluon energy. 

So far, the kernel $\omega(\vk)$, which defines the vacuum wave functional Eq.~(\ref{201}) 
and thus our basis of the Fock space, is completely arbitrary and, in principle, we could
use any positive-definite kernel $\omega(\vk)$. As long as we include the complete set
of states and do not introduce any approximation, the thermal expectation values will be
independent of $\omega(\vk)$. However, due to approximations necessary as, for example,
the restriction to two loops, the thermal averages will depend on the $\omega(\vk)$ chosen
and the optimal choice is obtained by extremizing the free energy Eq.~(\ref{278-13})
with respect to $\omega (\vk)$
\be\label{531-29}
\frac{\delta {\cF}}{\delta \omega (\vk)} = 0 ,
\ee
which yields the finite temperature gap equation
\be
\label{536-30}
\omega (\vk) = \vk^2 + \chi^2 (\vk) + I^{(0)} + I (\vk) ,
\ee
where
\begin{align}
I^{(0)} &= \frac{g^2 N_\mathrm{c}}{4} \int \dbar{q} \: \frac{3 - (\hat{\vk}\cdot\hat{\vq})^2}{\omega (\vq)} \, [1 + 2n (\vq)] , \nonumber \displaybreak[1]\\
I(\vk) &= \frac{g^2 N_\mathrm{c}}{4} \int \dbar{q} \: \bar{F}(\vk -\vq) \,
\frac{1 + (\hat{\vk} \cdot\hat{\vq})^2}{\omega (\vq)} \nonumber\\
&{}\Bigl\{ \left[ 1 + 2n (\vq) \right] \left[ \omega^2 (\vq) - \omega^2 (\vk) +
\bigl( \chi(\vq) - \chi(\vk) \bigr)^2 \right] \nonumber \\
& - 2 \omega (\vq) \bigl[ \chi(\vq) - \chi(\vk) \bigr]
\Bigr\} .\label{119}
\end{align}
These loop integrals, as well as the ones of the curvature [Eq.~\eqref{490-25}] and of the ghost Dyson-Schwinger
equation [Eq.~\eqref{501-27}], are ultraviolet divergent and need to be regularized and eventually renormalized.

Inserting the explicit expressions Eqs.~(\ref{482-24}) for the energy densities $e$ into
Eq.~(\ref{518-28}) and using the gap equation (\ref{496-26}), one finds
\be\label{660}
\begin{split}
\Omega (\vk) &= \omega (\vk) \bigl[ 1 + I_\Omega (\vk) \bigr] , \\
I_\Omega (\vk) &= \frac{g^2 N_\mathrm{c}}{4} \int \dbar{q} \: \bar{F}(\vk - \vq) \:
\frac{1 + (\hat{\vk} \cdot \hat{\vq})^2}{\omega (\vq)} \: \bigl[1 + 2n (\vq)\bigr] .
\end{split}
\ee

To carry out the renormalization we will have to deal with both finite- and zero-temperature
solutions. To avoid confusion, in the following we will explicitly indicate the temperature
dependence by writing $\omega(\vk, T)$, $d(\vk,T)$, \ldots\ instead of $\omega(\vk)$,
$d(\vk)$, \ldots

At very large momenta $\lvert\vk\rvert \gg T$ the temperature should become irrelevant.
Indeed, the temperature dependence of the loop integrals (which is due to the finite-temperature
occupation numbers $n(\vk)$) does not give rise to additional UV singularities.
Therefore the zero-temperature counterterms are sufficient to eliminate all UV singularities.

Adding the zero-temperature counterterms, see Ref.~\cite{Epple:2007ut}, and carrying
out the renormalization as described in Ref.~\cite{Reinhardt:2007wh}, one arrives at the
following renormalized gap equation
\be\begin{split}\label{584-31}
\omega^2 (\vk, T)  &= \vk^2 + \bar{\chi}^2 (\vk, T) + \Delta I^{(2)} (\vk, T) + c_0 + \bar{I}^{(0)}(T) \\
&+ 2 \bar{\chi}(\vk, T) \bigl[ \Delta I^{(1)} (\vk, T) + c_1 \bigr] + \bar I(\vk,T) ,
\end{split}
\ee
where we have introduced the abbreviations
\be\label{590-32}
\begin{split}
\bar{\chi} (\vk, T) &= \chi (\vk, T) - \chi (\mu_\chi, T = 0), \\
\bar{I}^{(0)} (T) &= I^{(0)}(T) - I^{(0)}(T = 0), \\
\Delta I^{(l)} (\vk, T) &= I^{(l)} (\vk, T) - I^{(l)} (\mu_\omega, T = 0) ,
\end{split}
\ee
and defined the following loop integrals
\begin{widetext}
\begin{align}
\label{29-186}
I^{(l)} (\vk,T) &= \frac{g^2 N_\mathrm{c}}{4} \int \dbar{q} \:
\bar{F}(\vk-\vq,T) \: \frac{1 + (\hat{\vk} \cdot \hat{\vq})^2}{\omega(\vq, T)}
\Bigl\{ \bigl[ \omega (\vq,T) - \bar{\chi}(\vq,T) \bigr]^l -
\bigl[ \omega (\vk,T) - \bar{\chi}(\vk,T) \bigr]^l \Bigr\} , \\
\bar{I} (\vk,T) &= \frac{g^2 N_\mathrm{c}}{4} \int \dbar{q} \,
\bar{F}(\vk-\vq,T) \: \frac{1 + (\hat{\vk} \cdot \hat{\vq})^2}{\omega(\vq,T)} \: 2 n(\vq)
\left\{ \omega^2 (\vq,T) - \omega^2 (\vk,T) +
\bigl[ \bar\chi(\vq,T) - \bar\chi(\vk,T) \bigr]^2 \right\} .
\end{align}
\end{widetext}

In addition, the renormalized equation (\ref{501-27}) for the ghost form factor reads 
\begin{equation}\label{d-renor-0}
\frac{1}{d(\vk,T)} = \left[ \frac{1}{d(\mu_d,T=0)} + I_d(\mu_d,T=0) \right] - I_d(\vk,T) .
\end{equation} 
The renormalized Dyson-Schwinger equations (\ref{584-31}) and \eqref{d-renor-0} contain
the finite renormalization scales, $\mu_i = \mu_\chi$,~$\mu_\omega$,~$\mu_d$, and the
renormalization constants $g(\mu_i)$, $\chi(\mu_i)$, $c_0(\mu_i)$, and $c_1(\mu_i)$.
The last two originate from the counterterms in the Hamiltonian and $\chi(\mu_i)$ from
the renormalization of the Faddeev-Popov determinant. In particular, in Ref.~\cite{Reinhardt:2007wh}
it was shown that for  $\mu_\omega = \mu_d = 0$ the value $c_1 = 0$ is required in order
that the 't Hooft loop obeys a perimeter law and is also favored by the variational principle. 
It was also found that the parameter $c_0$  has no influence on the IR- or UV-behavior of
the resulting solutions and influences only the mid-momentum regime of $\omega(\vk)$.
The choice of renormalization conditions for our study at finite temperature will be discussed
in Sec.~\ref{NR}. 

The Gribov-Zwanziger confinement scenario assumes that $d^{-1}(0,T=0)=0$. For practical
reasons, in the present paper we will assume a small but finite $d^{-1}(0,T)$, which results
in a massive gluon propagator, referred to as subcritical solution in Ref.~\cite{Epple:2007ut}.
This solution does not provide a confinig Coulomb potential, but for phenomenological
purposes may be as useful as the critical confining solution $d^{-1}(0)=0$ (see
Ref.~\cite{Epple:2007ut} for further discussions). One can give arguments that a $d^{-1}(0)\neq0$
is the result of an improper treatment of the Gribov problem \cite{Maas:2009se,Watson:2010cn}. In fact, it was explicitly
demonstrated in $1+1$ dimensions \cite{Reinhardt:2008ij}, 
and also arguments were given for $3+1$ lattice gauge theory in Landau gauge \cite{Maas:2009se},
that extending the functional integral over the transverse
gauge field to higher Gribov regions reduces the infrared strength of the ghost form factor,
pushing $d^{-1}(0)$ to higher values.
Based on this observation, it was argued in Refs.\ \cite{Watson:2010cn}
and \cite{Maas:2009se} that choosing different values of $d^{-1}(0)$ corresponds to
different ``gauge fixings''. (After all, a complete gauge fixing implies also the restriction
to the fundamental modular region, which is a subset of the first Gribov region.) Presumably,
in more than $1+1$ dimensions the restriction to the fundamental modular region requires
$d^{-1}(0)=0$. In any case this value is required for a linearly rising Coulomb potential,
which is a necessary condition for confinement in the Gribov-Zwanziger confinement scenario
\cite{Gribov:1977wm,Zwanziger:1993dh}.
Thus, if $d^{-1}(0)$ is kept finite for technical reasons,
it has to be kept small to stay close to the physical confining limit
$d^{-1}(0)=0$.


\section{\label{NR}Numerical results}

As shown in Sec.~\ref{subsec:proj}, the color projection removes
the zero mode from the Coulomb potential, see Eq.~(\ref{525-8}).
In the continuum it is replaced by
\begin{equation}\label{coulreg}
g^2 \bar{F}(\vk) = \lim_{\epsilon\to0} \frac{d^2(\vk)}{\vk^2 + \epsilon^2} \:\bigl[ 1 - \exp(-\vk^2/\epsilon^2) \bigr],
\end{equation} 
were we used the approximation [cf.~Eq.~(\ref{414-G4})] \cite{Feuchter:2004mk} 
\begin{multline}\label{appr} 
\langle F^{ab}_A(\vx, \vy) \rangle  = \sum_c \int \d[3]{z}
\langle \langle \vx, a \rvert (- \vec{D} \vec{\nabla})^{- 1}  \lvert c,\vz \rangle  \rangle  \\
\times  (- \vec{\nabla}^2)_{\vz}
  \langle \langle \vz, c \rvert (- \vec{D} \vec{\nabla})^{- 1}  \lvert \vy,b \rangle  \rangle ,
\end{multline}
with the external $\langle \cdots \rangle$ referring to the thermal average.
In the confining limit 
\begin{equation}
\label{conf}
g^2 F(\vk) = \frac{d^2(\vk)}{\vk^2} \xrightarrow{\vk \to 0} \frac{8\pi \sigma_{\textsc{c}}}{k^4} 
\end{equation} 
and the single, quasi-gluon energy in Eq.~(\ref{660}) is infinite at all temperatures,
which is certainly an artifact of our approximation, since
at least for large $\lvert\vk\rvert$ the quasi-gluon energy should be finite due to
asymptotic freedom.
For infinite
$\Omega(\vk)$ the finite-temperature occupation numbers $n(\vk)$ [Eq.~\eqref{372}] vanish
at all temperatures and there is no finite-temperature phase transition.
Thus the presently used approximations are inappropriate for the strictly
confining solution.
\begin{figure}[t]
\includegraphics[width=.9\linewidth]{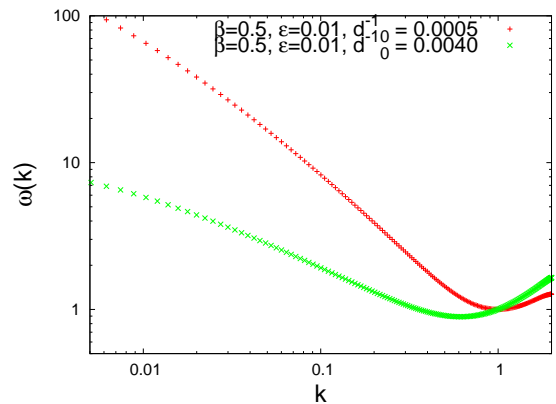}
\caption{\label{om-T=0.5-eps=0.01-d=0.0005}Low momentum (IR) behavior for $\beta=0.5$,
  $\epsilon = 0.01$, and $d^{-1}_0  = 0.0005,0.0040$
  of solutions for gap equation for $\omega$. In the limit
  $\epsilon \to 0$ and/or $\beta \to \infty$ the solutions do not
  change qualitatively. The critical solution corresponds to $d_0^{-1} = 0$, and the solution
  with $d_0^{-1} = 0.0005$ is close to critical. The IR limit is weakened as $d_0^{-1}$
  increases and the ghost propagator becomes massive.}
\end{figure}
\begin{figure}[t]
\includegraphics[width=.9\linewidth]{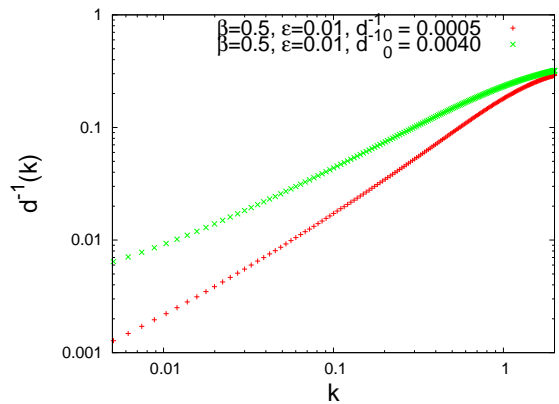}
\caption{\label{d-T=0.5-eps=0.01-d=0.0005}Same as Fig.~\ref{om-T=0.5-eps=0.01-d=0.0005} for the ghost form factor $d$.}
\end{figure}
As discussed in Ref.~\cite{Epple:2007ut}, without the approximation of Eq.~(\ref{appr})
there are no strictly confining solutions in the sense of Eq.~\eqref{conf} in the
variational approximation, and in this case the $\Omega(\vk)$ are finite (for finite $\vk$)
and at finite temperature a non-trivial solution with 
$n(\vk) \ne 0$ is expected. In this case we solve the set of finite-temperature Dyson-Schwinger
equations numerically on a momentum grid. The non-confining solutions depend on the
renormalized coupling or, alternatively, the value of $d(\vk=0,T)$  [cf.~Eq.~(\ref{501-27})].
As discussed in Sec.~\ref{VP}, the Dyson-Schwinger equations are renormalized by
subtraction at zero temperature to account for temperature-independent counterterms. In
the numerical computation, however, it is very difficult to solve these equations at
fixed $T$ unless subtracted at the same value of $T$. Thus, in the numerical results that
follow all subtractions will be done at finite $T$. In particular, when solving for
$d(\vk,T)$ [see Eq.~(\ref{501-27})] we use  
\begin{equation} 
\frac{1}{d(\vk,T)}  =  \left[  \frac{1}{d(\mu_d,T)}  + I_d(\mu_d,T) \right]  - I_d(\vk,T) .
\end{equation}
Comparing this with Eq.~(\ref{d-renor-0}) we have
\begin{equation} 
\label{renorcond} 
d_0^{-1} \equiv \frac{1}{d(\mu_d,T)}  =  \frac{1}{d(\mu_d,0)} + I_d(\mu_d,0) - I_d(\mu_d,T) ,
\end{equation} 
with $\mu_d$ chosen to be the lowest point on the  momentum grid, which
corresponds to $\mu_d =0$ in the infinite volume limit. 
In other words, we fix $d_0^{-1}$ with the temperature: At each temperature
we thus control the distance to the confining limit of the color Coulomb potential.
This implies that the mass scale which enters $I_d(\mu_d\sim 0,T)$ on the r.h.s.\ of Eq.~(\ref{renorcond}) depends on $T$. 
Similarly, the numerical stability of the solution of the gap equation for $\omega(\vk,T)$ [Eq.~(\ref{584-31})] 
requires that we use temperature-dependent renormalization constants, i.e., in Eq.~(\ref{590-32})
instead of subtracting at $T=0$ we subtract at finite $T$, so that $\mu_\chi = \mu_\chi(T)$ and $\mu_\omega =  \mu_\omega(T)$.
In particular, we use a single renormalization scale and set $\mu(T) = \mu_\omega(T) = \mu_\chi(T)$. 
This implies that we renormalize the gap equation at a finite momentum $\mu_\omega \ne 0$.
By renormalizing at $\mu_\omega = 0$, one would be enforcing a particular IR limit of the
solution of the gap equation, which could turn out to be incompatible with the finite-temperature
equation. Instead, by choosing $\mu_\omega$ away from the IR limit the value obtained from
solving numerically for $\omega(0) \equiv \omega(\vk=0,T)$ will serve to illustrate
the onset of a phase transition.

\begin{figure}
\includegraphics[width=.9\linewidth]{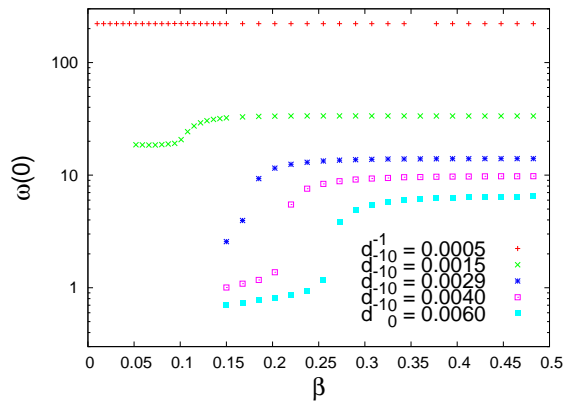}
\caption{\label{T=0.5-0.15-eps=0.01-phase}$\omega(0)$ as a function of temperature for $\epsilon=0.01$ and
  $d^{-1}_0 = \{0.0060$, $0.0040$, $0.0029$, $0.0015$, $0.0005\}$. The phase transition is clearly visible
  and becomes stronger and moves to lower temperatures as $d^{-1}_0$ increases.
  For $d_0^{-1} \to 0$ the phase transition disappears, i.e.,  moves to infinite temperatures.} 
\end{figure}

\begin{figure}[t!]
\includegraphics[width=.9\linewidth]{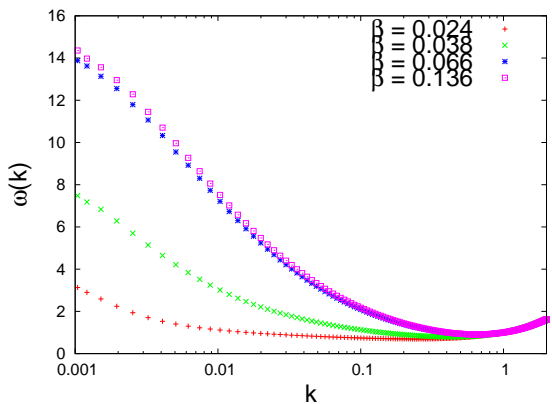}
\caption{Gap function $\omega(\vk)$ as a function of temperature ($d_0^{-1} = 0.0029$, 
$\epsilon=0$).
As the temperature increases $\omega(\vk)$ becomes less IR enhanced.}
 \label{omega-full}
\end{figure}

To search for the phase transition we proceed as follows. We start with a small but finite
IR regulator $\epsilon$ [see Eq.~\eqref{coulreg}], and at given, small $T$ (large $\beta$)
we choose a solution close to a critical one. A typical case is shown in
Figs.~\ref{om-T=0.5-eps=0.01-d=0.0005} and \ref{d-T=0.5-eps=0.01-d=0.0005}.
In all figures, physical, dimensional quantities are plotted in units of $\mu(T)$. 
With fixed $\epsilon$ and $d_0$ we increase $T$ (decrease $\beta$) and study both $\omega$ and $d$. 
A series of computations of $\omega(0)$ as function of $\beta$ for $\epsilon=0.01$ and
$d_0^{-1}$ in the range [0.0005--0.0060]
 is shown in Fig.~\ref{T=0.5-0.15-eps=0.01-phase}. 
As $d_0^{-1}$ increases, the solution becomes less critical, i.e., less IR enhanced, and
the Coulomb potential moves away from the confining limit of Eq.~(\ref{conf}). In this phase,
as shown in Fig.~\ref{T=0.5-0.15-eps=0.01-phase}, there is an abrupt change in the gap function
$\omega(0)$ at a critical temperature which decreases as the solution become weaker in the IR.
Please note that the results shown for different values of $d^{-1}(0)$ correspond
to different physical scales.
\begin{figure}
\includegraphics[width=.9\linewidth]{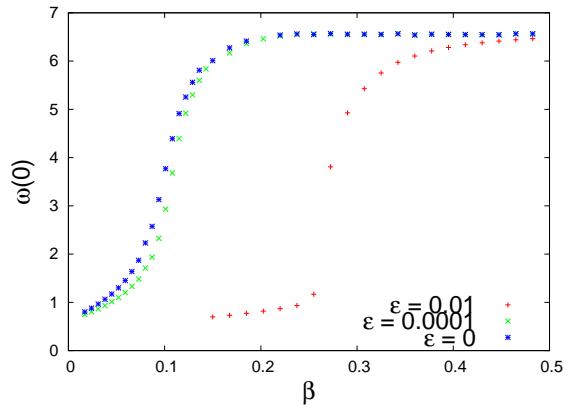}
\caption{\label{T=0.5-0.15-eps=0.0001-d=0.060}$\omega(0)$ as a function of temperature
  for $d_0^{-1} = 0.0060$ massive solution for $\epsilon=0.01$ (as in Fig.~\ref{T=0.5-0.15-eps=0.01-phase}) compared with 
  with solutions for $\epsilon=0.0001$ and $\epsilon=0$.  } 
\end{figure}
\begin{figure}
\includegraphics[width=.9\linewidth]{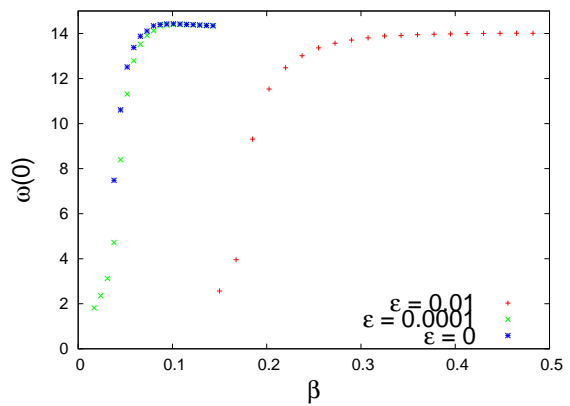}
\caption{\label{T=0.5-0.15-eps=0.0001-d=0.029}Same as in Fig.~\ref{T=0.5-0.15-eps=0.0001-d=0.060}
 for $d_0^{-1} = 0.0029$. } 
\end{figure} 
We also studied the dependence on $\epsilon$, as the limit $\epsilon\to0$
should be taken to approach the infinite volume. Starting from a massive solution at zero temperature,
e.g., with $d_0^{-1} = 0.060$, we increase the temperature and decrease $\epsilon$. In particular,
we solve the equations for $\epsilon=0.0001$ and $\epsilon=0$. The latter choice is possible,
since with an IR finite ghost dressing function ($d_0^{-1} \ne 0$)
the Coulomb potential in Eq.~(\ref{conf}) does not lead to an IR
singularity in the integrals. The results are shown in Figs.~\ref{T=0.5-0.15-eps=0.0001-d=0.060} and
\ref{T=0.5-0.15-eps=0.0001-d=0.029}.
Even though, in numerical simulations with a finite momentum grid,
one cannot reach the exact critical limit of $\epsilon = 0$, $d_0^{-1} = 0$ (which implies
$F(\vk) \propto 1/k^4$ at zero temperature), the numerical results shown in these figures 
 are consistent with the anticipated (see Sec.~\ref{intro}) disappearance of the phase transition in the Coulomb phase.
In particular, we observe that, as $d_0^{-1}$ decreases,
the gap function $\omega(0)$ grows and eventually becomes infinite as it is the case in
the zero-temperature limit. In other words, the phase transition moves to
infinite temperature ($\beta \to 0$) as the gluon self-energy
becomes infinite. In Fig.~\ref{omega-full} we show the evolution with temperature of $\omega(\vk)$ as a function of gluon momentum. 
As expected, the gap function becomes less IR enhanced as temperature increases ($\beta$ decreases). 


\section{\label{SO} Summary and Outlook} 

We studied the temperature dependence of QCD correlation functions with a variational
ansatz for the gluon density matrix in the Coulomb gauge. 
The resulting one-loop Dyson-Schwinger equations for the gluon propagator
and ghost form factor $d(\vk)$ were solved numerically, assuming a subcritical behavior,
i.e., $d^{-1}(0)\neq0$, which, however, was chosen close to the critical one, $d^{-1}(0)=0.$

Strictly speaking, the variational Coulomb gauge model which leads to $d_0^{-1} \ne 0$ is 
not confining and thus only loosely related to QCD. The Gribov-Zwanzinger confinement
scenario is reached in the $d_0^{-1} \to 0 $ (and $\epsilon \to 0$) limit. We find it amusing, however, that 
the quasi-gluons which for $d_0^{-1} \ne 0$ are  deconfined at all temperatures 
behave similar to the physical gluons both below and above $T_c$.   
We have found that quasi-particle, gluonic excitations built on top of such a subcritical vacuum
lead to a sharp transition in the above correlations functions. To solve the Dyson-Schwinger equations
we used temperature dependent renormalization conditions. This results in the phase transition point
in Fig.~\ref{T=0.5-0.15-eps=0.01-phase}  moving  with $d_0^{-1}$. By fixing the critical temperature
to a physical value, if known, this variation could be used to determine the function $\mu(T)$  and,
ultimately, the temperature dependence of these correlation functions. As one tunes the zero-temperature
solution to approach the critical limit, $d_0^{-1} \to 0$, the phase transition moves
to infinite temperatures. 
Even though thermal excitations are restricted to color single states, contribution to the partition function from two 
gluons is $O(1/V)$  compared to that of a glueball. The former are thus expected to make
negligible contribution in the thermodynamical limit, which explains why a confining Coulomb potential
at zero temperature remains confining at finite temperatures~\cite{Greensite:2004ke}.
However, the confining potential 
can bind gluons into color-singlet glueballs and a phase transition could be observed, for example in a change of the
radius of the glueball wave function. We will consider such a mixed glueball/quasi-gluon phase in the forthcoming work. 
   
\begin{acknowledgments}
H.R.\ and D.R.C.\ are grateful to P.~Watson for a critical reading of
the manuscript and useful comments.
H.R.\ and D.R.C.\ have been supported by the Deu\-tsche For\-schungsgemeinschaft (DFG) under contract
No.\ DFG-Re856-3 and by the Cusanuswerk--Bisch\"ofliche Studienf\"orderung.
A.P.S.\ research is supported in part by the U.S.\ Department of Energy under Grant No.~DE-FG0287ER40365.
\end{acknowledgments}

\end{document}